\documentclass{osa-article}

\journal{oe}
\usepackage{siunitx}

\begin{document}

\title{Third-harmonic light polarization control in magnetically-resonant silicon metasurfaces}

\author{Andrea Tognazzi,\authormark{1,2} Kirill I. Okhlopkov,\authormark{3} Attilio Zilli, \authormark{4} Davide Rocco,\authormark{1,2} Luca Fagiani,\authormark{4,5} Erfan Mafakheri,\authormark{5} Monica Bollani,\authormark{5} Marco Finazzi, \authormark{4} Michele Celebrano, \authormark{4} Maxim R. Shcherbakov,\authormark{3} Andrey A. Fedyanin,\authormark{3} and Costantino De Angelis\authormark{1,2,*}}

\address{\authormark{1}Department of Information Engineering, University of Brescia, Via Branze 38, 25123 Brescia, Italy\\
\authormark{2}CNR-INO (National Institute of Optics), Via Branze 45, 25123 Brescia, Italy\\
\authormark{3}Faculty of Physics, Lomonosov Moscow State University, Moscow 119991, Russia\\
\authormark{4}Department of Physics, Politecnico di Milano, Piazza Leonardo Da Vinci 32, 20133 Milano, Italy\\
\authormark{5}CNR-IFN, LNESS laboratory, Via Anzani 42, 22100 Como, Italy}

\email{\authormark{*}costantino.deangelis@unibs.it} 
\begin{abstract}
Nonlinear metasurfaces have become prominent tools for controlling and engineering light at the nanoscale. Usually, the polarization of the total generated third harmonic is studied. However, diffraction orders may present different polarizations. Here, we design an high quality factor silicon metasurface for third harmonic generation and perform back focal plane imaging of the diffraction orders, which present a rich variety of polarization states. Our results demonstrate the possibility of tailoring the polarization of the generated nonlinear diffraction orders paving the way to a higher degree of wavefront control.
\end{abstract}

\section{Introduction}
Manipulation of light is of paramount importance in many fields such as opto-electronics, image processing, sensing and cryptography\cite{bib:Ma2018,bib:Liu2011}. The 2D nature of metasurfaces, which are composed by an array of resonators, makes them suitable candidates for compact photonic devices\cite{bib:Kim2018,bib:Scheuer2020}. The optical properties of such structures can be tailored by tuning the geometrical parameters of each resonator in the periodic array or by changing the material\cite{bib:Minovich,bib:He2019,bib:Seyedeh2018}. Thanks to the improved accuracy of nanofabrication techniques, it is nowadays possible to obtain high quality nano-objects consisting of metals or dielectric materials to manipulate light in the visible and near-infrared regimes\cite{bib:Naffouti2017}. The applications of metasurfaces include beam steering \cite{bib:Wei2013,bib:Forouzmand2016}, light focusing\cite{bib:Schlickriede2020,bib:Benali2020}, holography\cite{bib:Hu2019}, and sensing\cite{bib:Ferry2020}.

Implementing nonlinear optics at the nanoscale is very challenging because one cannot exploit phase matching, which can be achieved only over mesoscopic scales. In this frame, the added value of metasurfaces consists in the possibility of exploiting collective modes stemming from the interactions between neighboring nanoresonators to enhance the local electric field, improve the conversion efficiency, and tailor the emitted radiation\cite{bib:Shcherbakov2019,bib:Li2017,bib:Ningning2015,bib:Sun2017}.
The low losses make dielectrics more suitable than metals for second- and third-harmonic generation (THG), all-optical switching and modulation of visible and near-infrared light\cite{bib:Carletti2017,bib:Rocco2018,bib:Staude2013,bib:Gili2018,bib:Morales2016,bib:Marino2019,Shcherbakov2015,Shcherbakov2017}. In the past few years, high-refractive index dielectric materials were employed to build nanoresonators to improve nonlinear frequency conversion \cite{bib:Liu2018,Shcherbakov2014} and manipulate light emission \cite{bib:Ghirardini2018,bib:Carletti2018}. One of the most attractive materials for nanophotonics applications is silicon due to its well-established fabrication technology, high-refractive index and technological relevance\cite{bib:Staude2017,bib:Kyu2019}. Previously, nonlinear beam deflection has been achieved by inducing a phase shift using different building blocks\cite{bib:Wang2018}. However, the polarization of the diffraction orders is usually an overlooked property when studying nonlinear gratings \cite{Loechner2018}. In this article, we report the design and fabrication of high quality factor ($Q$-factor) metasurfaces and we propose a simple electromagnetic model to explain the polarization of the third harmonic (TH) diffraction orders. Orthogonal polarizations are measured for different diffraction orders depending on the dominant multipolar component at resonance. Our results pave the way to the realization of a higher degree of polarization-controlled nonlinear diffractive metasurfaces.

\begin{figure}
\centering
\includegraphics[width=0.5\columnwidth]{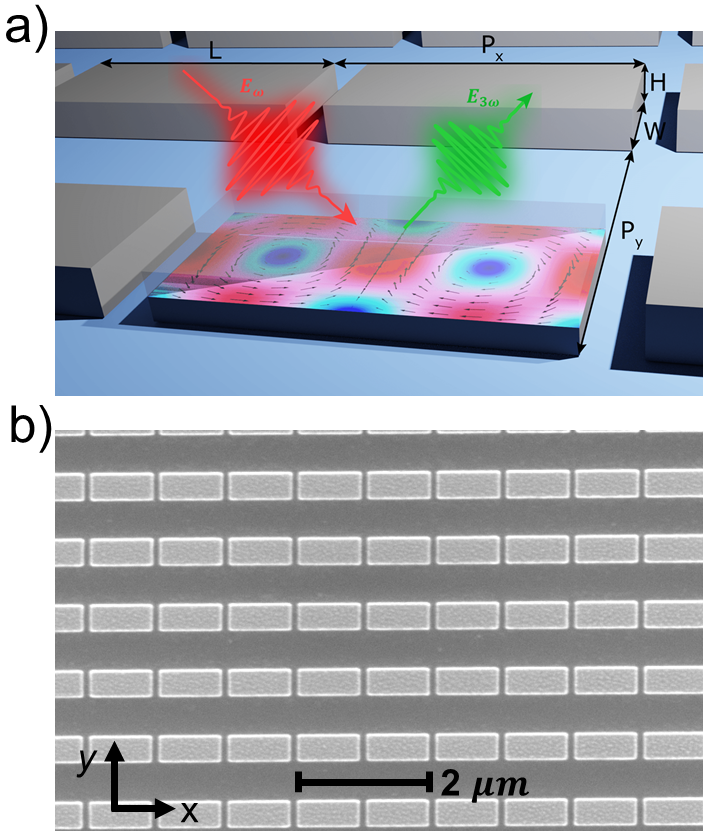}
  \caption{(a) Sketch of the metasurface and of the electric field distribution at normal incidence showing the magnetic quadrupole behaviour of the fundamental frequency. The near field distribution can be used to predict the diffraction orders polarization. The unitary cell is constituted by a silicon cuboid laying on a SiO$_2$ substrate. (b) SEM planar view image of a dielectric metasurface image.}
  \label{fig1}
  \end{figure}

\begin{figure}
\centering
\includegraphics[width=0.7\columnwidth]{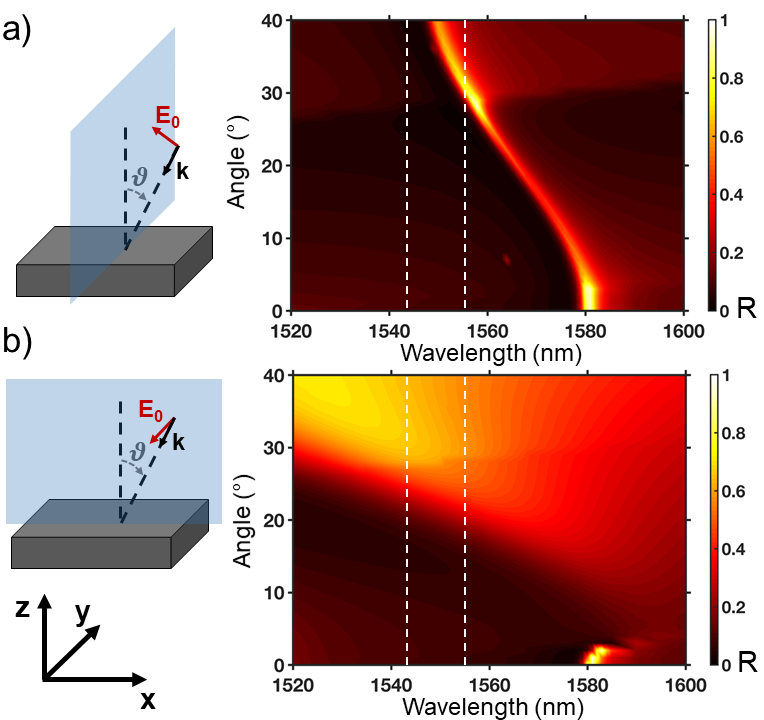}
  \caption{Simulated reflectance for $p$ (a) and $s$ (b) incident polarization as a function of the wavelength and the incidence angle $\vartheta$. The dashed white lines delimit the FWHM bandwidth ($1554 \pm 8$~\si{nm}) of the pump used in the experiments. For incident \textit{p}-polarization, the high quality factor is preserved when $\vartheta$ increases and the resonance blue shifts. For incident \textit{s}-polarization, the high quality factor resonance fades away when $\vartheta$ increases and a broader resonance appears at shorter wavelengths.}
  \label{fig2}
\end{figure}

\section{Design and fabrication}
We employ a commercial finite element solver (Comsol Multiphysics) to optimize the design of high-$Q$ metasurfaces made of silicon cuboids arranged in a periodic rectangular lattice (see Fig.~\ref{fig1}a). 
We created a waveguide-like system with a channel coupling the light and the structure to obtain a metasurface with different quality factors depending on the excitation geometry. In the experiments, we achieve the resonant condition by changing the angle of incidence, this allows us to excite two different modes under incident \textit{p}- and \textit{s}-polarization.
The metasurfaces are realized on a silicon-on-insulator (SOI) substrate with a device layer of $H=125$~\si{nm} on 2~\si{\micro m} of buried oxide (see Fig.~\ref{fig1}b). Arrays of rectangles (width $W=428$~\si{nm}, length $L=942$~\si{nm} and periodicity $P_x$ and $P_y$ 1065--1060~\si{nm}, respectively) aligned along the [110] direction are patterned by means of e-beam lithography (EBL) and reactive ion etching (RIE). The resist is spin-coated on the SOI substrate and then exposed to the electron beam of a converted scanning electron microscope (SEM) along the designed pattern (acceleration voltage of 30~\si{kV}). A double layer of PMMA diluted in chlorobenzene, respectively, at 3.5\% and 1.5\% is employed. The dose used for the structures is 350~\si{\micro C/cm\textsuperscript{2}}. After exposure, PMMA is developed in a solution of methyl isobutyl ketone (MIBK) and isopropanol (IPA) in a 1:3 ratio; MIBK is diluted in order to obtain well-defined profiles. The sample is immersed in this solution and agitated manually for 90~\si{s}; a pure IPA solution is used for 1 minute to stop the development of the resist. Then, the pattern is transferred to the thin Si film by RIE in a CF$_4$ plasma, using 80~\si{W} of radio frequency power and a total gas pressure of 5.4~\si{mTorr}. Finally, the resist is removed using acetone and the sample surface is exposed to O$_2$ plasma in order to remove any residual resist.
In the simulations, we model the silicon refractive index as reported in \cite{bib:Green2008} and we assume a wavelength-independent refractive index (\textit{n}=1.45) for the SiO$_2$ substrate. The spatial period $P_x=P_y=P$ of the through-notches has been chosen to satisfy the matching condition resulting from momentum conservation between a normal incident plane wave with in-plane modes, i.e. $2\pi/P=\beta(\omega)$ where $\beta(\omega)$ is the propagation wavevector of the mode\cite{bib:Shcherbakov2019}. Possible deviations of $\beta(\omega)$ in the fabricated device from the simulated value can be matched by tuning the wavelength $\lambda_0$ or the angle $\theta$ of the incident plane wave. Fig.~\ref{fig2} shows sketches of the incident polarization and the reflectance (\textit{R}) at the fundamental frequency (FF) as a function of $\lambda_0$ and $\theta$ for \textit{p} and \textit{s} polarized excitation with $\vec{E}_0(\omega) \perp \hat{x}$. When $\vec{E}_0(\omega)$ is \textit{p} polarized the metasurface shows a sharp resonance ($Q=399$) which blue-shifts when $\vartheta$ is varied, albeit maintaining a narrow spectral width. When the impinging wave is \textit{s} polarized, it excites a magnetic dipole mode with a lower $Q$ factor ($Q=29$). We then simulate the TH field $\vec{E}(3\omega)$ by evaluating in a second step of the computation the nonlinear current generated within the structures by the fundamental field $\vec{E}_0(\omega)$ through the third-order susceptibility as reported in \cite{bib:Smirnova2016}. The diffraction orders are calculated by performing the Fourier transform of the near field simulated at the TH frequency.

\section{Experiments}

\begin{figure}
\centering
\includegraphics[width=\columnwidth]{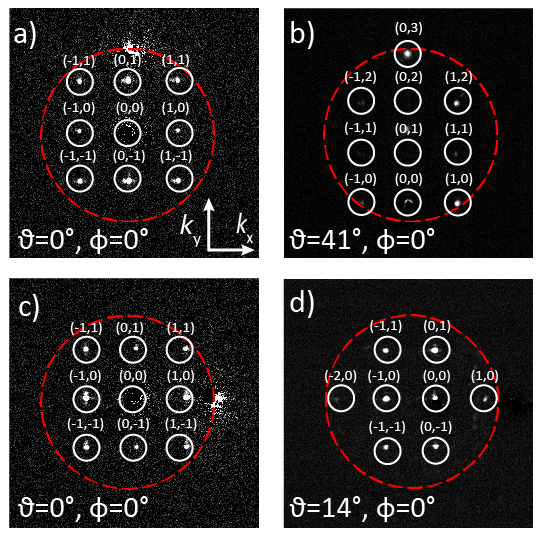}
  \caption{(a,c)~Experimental BFP images under normal incidence excitation. The red circle represents the numerical aperture of the objective (NA=0.85). (b,d) BFP images with tilted illumination at 41\textdegree~for \textit{p} and 14\textdegree~for \textit{s} polarized light. When the angle of incidence is increased, the (0,0) order is not emitted perpendicularly to the metasurface due to the in-plane components. For \textit{p}-polarization, the diffraction orders move along $k_y$ since the incident beam wavevector lies in the \textit{yz}-plane, while, for \textit{s}-polarization, they move along $k_x$ since the incident wavevector is in the \textit{xz}-plane. As a consequence, at certain angles, some diffraction orders disappear and others fall in the NA view window.}
  \label{BFP_angle}
\end{figure}

We employ a pulsed laser (160 fs) centered at 1554~\si{nm} (FWHM=17~\si{nm}) and focus the beam in the back focal plane (BFP) of a 60x objective (Nikon, CFI Plan Fluor 60XC, NA=0.85) to obtain a loosely focused beam on the sample. We shift the pump beam in the BFP plane of the objective to change the angle of incidence. We collect the emitted TH through the same objective used for the excitation, and chromatically filter it. A Bertrand lens in the detection path focuses the TH beam in the BFP of the objective to image the TH diffraction orders, while a polarizer is used to analyze the polarization of the TH. A cooled CCD camera is used to acquire BFP images such as the ones in Fig.~\ref{BFP_angle}, where $\vartheta$ is the angle of incidence and $\phi$ is the polarizer angle with respect to the \textit{x}-axis.

\section{Results}
\begin{figure}
\centering
\includegraphics[width=\columnwidth]{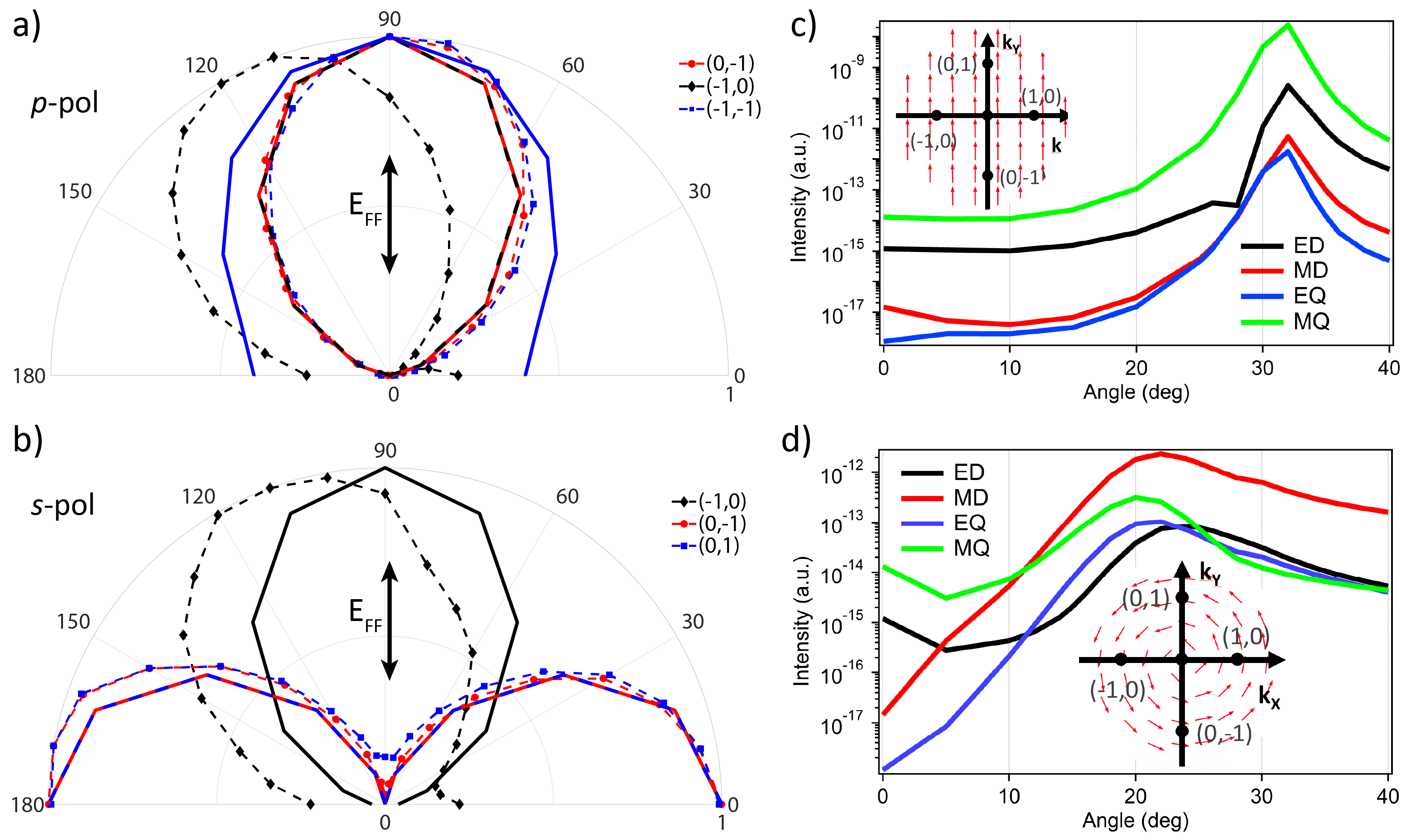}
  \caption{Experimental (dashed) and simulated (continuous) polarization-resolved TH power excited at the angle with maximum THG for $p$ (a) and $s$ (b) polarization, respectively. The vertical double arrows represent the incident pump polarization. (c,d) Cartesian multipolar decomposition for incident $p$ and $s$ polarization, respectively. The insets in (c,d) represent the far field polarization of the magnetic quadrupole and of the magnetic dipole, respectively.}
  \label{fig:polarization}
\end{figure}

In Figs.~\ref{fig:polarization}(a,b) the intensity of TH light as function of the analyzer angle of different diffraction orders is reported for experiments (dashed) and simulations (continuous) for \textit{s} and \textit{p}-polarized fundamental wavelength light illuminating the sample at the incidence angle leading to maximum THG. Normalized experimental data in Fig.~\ref{fig:polarization}a refer to \textit{p}-polarized light impinging on the sample at 41$^\circ$, corresponding to the maximum THG signal, while simulations corresponds to  32\textdegree. All the THG diffraction orders are polarized along the $y$ axis as predicted by the simulations. THG is maximum for incident \textit{s}-polarization at 14$^\circ$ in the experiment and 22$^\circ$ in the simulations.
The discrepancies in the resonant angles and the systematic tilt of the (-1,0) diffraction order polarization may be due to the uncertainty in the experimental pump beam angle of incidence and fabrication defects. In order to have a better insight on the polarization of diffraction orders we performed a cartesian multipolar decomposition of the TH field (see Fig.~\ref{fig:polarization}c,d). For incident \textit{p}-polarization, the main multipolar component is always a magnetic quadrupole, $Q_{xz}$, whose amplitude is maximum at the resonant angle, leading to no variations of the diffraction order polarization when the angle of incidence is changed . For \textit{s}-polarization, the main multipolar component changes at resonance and becomes a magnetic dipole along the \textit{z}-axis, with a spatially non uniform far field polarization (see Fig.\ref{fig:polarization}d inset). This corresponds to a variation of the polarization of all the diffraction orders $(m,n)$ with $n\neq 1$. The polarization of the diffraction orders can be described by a simple formula, which takes into account the electromagnetic field distribution of each scatterer and the periodic structure. In the far-field region the total electric field radiated ($E_t$) by the metasurface is proportional to the far field radiated by the single array element ($E_s$) through the array factor ($AF$):
\begin{equation}
    \vec{E}_t^{3\omega}=\vec{E}_s^{3\omega} AF(P,3\omega),
\end{equation}
where the $AF(P,3\omega)$ is a function that depends only on the periodicity of the array and the TH frequency. Here, each cuboid can be envisioned as an antenna whose emission pattern is determined by the superposition of all the multipolar components and the diffraction orders are determined according to the $AF$ as described in \cite{bib:Balanis}. This formalism enables one to tailor the polarization state of the nonlinear diffraction orders by engineering the main multipolar components at the TH frequency describing the single antenna behaviour. It is worth noting that this approach can be applied also for closely packed unit-cells once the multipolar decomposition is completely resolved for the meta-units that form the metasurface under test. 

\section{Conclusions}
We showed, both experimentally and numerically, a complex behaviour of the polarization of the TH diffraction orders as a function of the incidence angle of the fundamental pump beam. We applied a cartesian multipolar decomposition and a simple formula to describe the polarization of the diffraction orders and provide a method to tailor the far field properties of the metasurface. Our results demonstrate that the polarization of the  diffraction orders is solely influenced by the near field distribution within each single element, which can be modelled by considering the multipolar decomposition, while the far field is determined by the period of the metasurface relative to the exciting wavelength. Our description can be readily applied to second-harmonic generation and to any type of periodic metasurface.

\section*{Funding}
We acknowledge the financial support by the European Commission through the FET-OPEN projects "Narciso" (828890) and "METAFAST" (899673),  by the Italian Ministry of University and Research (MIUR) through the PRIN project “NOMEN” (2017MP7F8F) and by Russian Ministry of Science and Higher Education (No 14.W03.31.0008).

\section*{Disclosures}
The authors declare no conflicts of interest.

\bibliography{Silicon_metasurface}
\end{document}